\documentclass[prd,aps,preprint,tightenlines,showpacs,amsmath,amssymb]{revtex4}
\usepackage{graphicx}
\usepackage{feynmp}

\def\cancel#1#2{\ooalign{$\hfil#1\mkern1mu/\hfil$\crcr$#1#2$}}
\def\slash#1{\mathpalette\cancel{#1}}

\def\mpi2{m_\pi^2}
\def\mK2{m_K^2}

\newcommand{\bea}{\begin{eqnarray}}
\newcommand{\eea}{\end{eqnarray}}
\newcommand{\be}{\begin{equation}}
\newcommand{\ee}{\end{equation}}

\begin{document}
\bibliographystyle{apsrev}

\vspace*{-10mm}
\begin{flushright}
\normalsize
CU-TP-1166 \\
KANAZAWA-06-16\\
RBRC-623
\end{flushright}

\pacs{11.15.Ha, 
      11.30.Rd, 
      12.38.Bx, 
      12.39.Hg  
}


\title{Perturbative renormalization for static and domain-wall bilinears and four-fermion operators with improved gauge actions}

\author{Oleg~Loktik$^{1}$ 
         and
         Taku~Izubuchi$^{2,3}$
}

\affiliation{
$^1$Physics Department, Columbia University, New York, NY 10027, USA \\
$^2$RIKEN-BNL Research Center, Brookhaven National Laboratory, Upton, NY 11973, USA \\
$^3$Institute for Theoretical Physics, Kanazawa University, Kanazawa, Ishikawa 920-1192, Japan 
}

\date{December 19, 2006}

\begin{abstract}
We calculate one-loop renormalization factors for heavy-light bilinears as well as four-fermion operators relevant for $B^{0} - \bar{B}^{0}$ mixing calculations on the lattice. We use the static approximation for heavy quarks and the domain-wall formulation for light quarks. We present results for different choices of improved gauge action. 
\end{abstract}

\maketitle

\newpage


\section{Introduction}
\label{sec:intro}
Recent experimental observations \cite{Abazov:2006dm,Abulencia:2006mq,Abulencia:2006ze} intensified the interest in B-meson physics as the prime area in which the validity of the Standard Model can be put to a test. In order to perform such a test, theoretical predictions for $f_B$, $B_B$ and $\Delta m_s/\Delta m_d$ are needed. In principle, the lattice approach to QCD allows one to calculate these and other quantities of interest from the first principles. However, the majority of lattice results to date suffer from substantial systematic errors which make comparison with experimental results inconclusive. One of the largest systematic errors comes from the procedure which relates lattice operators to the corresponding operators in continuum QCD.

There are two main sources of error in this matching procedure. First, lattice actions used in simulations usually have a reduced symmetry group. The lack of symmetry leads to the introduction of additional lattice operators which complicates the matching. Second, the matching is usually performed at the one-loop level in perturbation theory. More often than not, the one-loop contributions turn out to be quite large. Large one-loop contributions undermine the perturbative approach to matching and force the introduction of a large systematic error in order to account for possible higher-order corrections.   

In this paper we will demonstrate that both of these issues can be addressed by an appropriate choice of lattice action. Specifically, we will focus on the combination of the static effective field theory \cite{Eichten:1989zv} for heavy quarks, the domain-wall formulation \cite{Kaplan:1992bt,Shamir:1993zy} for light quarks and an improved gauge action \cite{Luscher:1984xn} for gluons.  This action provides a good approximation for heavy mesons such as $B$, and it has already been employed in numerical simulations \cite{Gadiyak:2005ea}.

We will obtain the renormalization constants connecting bare lattice operators to continuum QCD operators renormalized in ${\rm \overline{MS}}$(NDR) scheme. In particular, we will consider heavy-light bilinears and four-fermion operators relevant for $B^{0} - \bar{B}^{0}$ mixing. The matching will be done at the one-loop level in perturbation theory. We will split the matching into two parts. First, the continuum QCD operators need to be related to the continuum static theory operators. This step has already been carried out, and we will quote the results here. Second, the continuum static theory operators need to be matched to their lattice counterparts. This second step has not been performed before for our choice of lattice action, and thus is the main subject of this paper.

This paper is organized as follows. In Sec.~\ref{sec:action} we introduce our choice of lattice action and discuss the corresponding Feynman rules. In Sec.~\ref{sec:symmetries} the symmetries of the action are discussed. In Sec.~\ref{sec:match_full_effective_op} we quote the results of matching the continuum QCD and continuum static theory operators. In Sec.~\ref{sec:match_effective_lattice_op} we will match the static theory bilinears to their lattice counterparts. In Sec.~\ref{sec:match_effective_lattice_4ferm_op} we present the matching of four-fermion operators as well as $B_B$. In Sec.~\ref{sec:mean_field_impr} we discuss the application of mean-field improvement to our results. 

\section{Lattice Action and Feynman Rules}
\label{sec:action}
Our lattice action consists of three parts:
\begin{equation}\label{eq:action}
S = S_{\rm static} + S_{\rm DW} + S_{\rm gauge} ,
\end{equation}
where $S_{\rm static}$ is the static-quark action describing heavy quarks, $S_{\rm DW}$ is the domain-wall fermion action describing light quarks and $S_{\rm gauge}$ is the pure gauge action. In this and all subsequent sections we set the lattice spacing $a$ equal to 1. 

                                                                                
\subsection{Static-quark action}
\label{subsec:hqet}
The simplest formulation of the static-effective-field-theory action on the lattice is given by \cite{Eichten:1989kb} 
\begin{equation}\label{eq:hqa} S_{\rm static} = \sum_{x} \bar{h}(x) [ h(x) - U_{0}^{\dagger}(x-\hat0)h(x-\hat0)], \end{equation} 
where $U_{0}(x-\hat0)$ is the gauge link in temporal direction between sites $x$ and $x-\hat0$. The static-quark field $h$ satisfies $\gamma_0 h = h$. The static-quark propagator in momentum space is given by
\begin{equation}
H(k) = \frac{1}{1- e^{-ik_0} + \epsilon}.
\end{equation}
The quark-gluon vertices are
\begin{equation}
\tilde{V}^a_\mu(k,k^{\prime}) = -igT^a\delta_{0\mu}e^{-i(k_0+k^\prime_0)/2},
\end{equation}
\begin{equation}
\tilde{V}^{ab}_{\mu\nu}(k,k^{\prime}) = -\frac{1}{2}g^2\{T^a,T^b\}\delta_{0\mu}\delta_{0\nu}e^{-i(k_0+k^\prime_0)/2},
\end{equation}
with $T_{a}$ being a generator of the ${\rm SU}(N_c)$ color group. We note that it is not difficult to extend the results of this paper to the case of the modified static-quark action of Ref.~\cite{DellaMorte:2003mn}. The results of such extension will be reported elsewhere.

\subsection{Domain-wall fermion action}
\label{subsec:dwf} 
For a quark with mass $m$ the domain-wall fermion action\cite{Shamir:1993zy,Furman:1994ky} is given by
\begin{equation}\label{eq:dwf_action}
S_{\rm DW} = \sum_{s,s^\prime=1}^{N} \sum_{x,y}\overline\psi_{s}(x) D^{\rm DW}_{ss^\prime}(x,y)\psi_{s^\prime}(y) + \sum_{x}m\overline q(x)q(x),
\end{equation}
\begin{equation}
D^{\rm DW}_{ss^\prime}(x,y) = D^4(x,y)\delta_{ss^\prime} + D^5(s,s^\prime)\delta_{xy} + (M_5 - 5)\delta_{ss^\prime}\delta_{xy},
\end{equation}
\begin{equation}
D^4(x,y) = \sum_\mu \frac{1}{2}\left[(1+\gamma_\mu)U_\mu(x)\delta_{x+\hat{\mu},y} + (1-\gamma_\mu)U^{\dagger}_\mu(y)\delta_{x-\hat{\mu},y}\right],
\end{equation}
\begin{equation}
D^5(s,s^\prime) = \left\{
\begin{array}{ll}
 P_R\delta_{2,s^\prime} & (s=1) \\
 P_R\delta_{s+1,s^\prime} + P_L\delta_{s-1,s^\prime} & (1<s<N) \\
 P_L\delta_{N-1,s^\prime} & (s = N) \\
\end{array}
\right.,
\end{equation}
where $\psi_{s}(x)$ is a (4+1)-dimensional Wilson-style fermion field. The fifth dimension extends from $1$ to $N$ and is labeled by $s$. $P_L = (1-\gamma_{5})/2$ and $P_R = (1+\gamma_{5})/2$ are the projectors for the left- and right-handed spinors. The domain-wall height $M_{5}$ is a parameter of the theory, which can be set $0 \leqslant M_5 \leqslant 2$. Finally, $q(x)$ represents the physical four-dimensional quark field constructed from the five-dimensional field $\psi_s(x)$ at $s=1$ and $N$
\begin{equation}\label{eq:phys_quark}
q(x) = P_{R}\psi_{1}(x) + P_{L}\psi_{N}(x) ,
\end{equation}  
\begin{equation}\label{eq:phys_antiquark}
\overline{q}(x) = \overline{\psi}_{1}(x)P_L + \overline{\psi}_{N}(x)P_R .
\end{equation}  
In momentum space the free field domain-wall Dirac operator is given by
\begin{equation}
D(p)_{st} = \sum_\mu i\gamma_\mu\sin p_\mu \delta_{st}  + (W^+(p)_{st} + m\delta_{s,1}\delta_{N,t})P_R + (W^-(p)_{st} + m\delta_{s,N}\delta_{1,t})P_L,
\end{equation}
where $s$ and $t$ label coordinates in the fifth dimension and 
\begin{equation}
W^+(p)_{st} = \left( \begin{array}{cccc}
-W(p) &  1      &       &       \\
      & \ddots   & \ddots &       \\
      &         & -W(p)   & 1     \\
      &         &       & -W(p)  \end{array} \right),
\end{equation}
\begin{equation}
W^-(p)_{st} = \left( \begin{array}{cccc}
-W(p) &        &       &       \\
 1    & -W(p)  &       &     \\
      & \ddots   & \ddots &       \\
      &         & 1      & -W(p)  \end{array} \right),
\end{equation}
\begin{equation}
W(p) = 1 -M_5 + \sum_\mu(1-\cos p_\mu).
\end{equation}
From now on we will focus on $m=0$ case and assume that $N$ is large. We will rely on the formalism developed in Refs.~\cite{Aoki:1997xg,Aoki:1998vv} in our perturbative treatment of domain-wall fermions. By inverting the Dirac equation one can write the tree-level DWF propagator as 
\begin{equation}\label{eq:dwf_prop}
S(p)_{st} = (-i\gamma_\mu \sin p_\mu \delta_{su} + W^-_{su})G^R_{ut}(p)P_R \\
+ (-i\gamma_\mu \sin p_\mu \delta_{su} + W^+_{su})G^L_{ut}(p)P_L, \end{equation}
where 
\begin{equation}
G^{R}_{st}(p) = -\frac{A}{F}[(1-We^{-\alpha})e^{-\alpha(2N-s-t)} 
+(1-We^{\alpha})e^{-\alpha(s+t)} ] + Ae^{-\alpha|s-t|},
\end{equation}
\begin{equation}
G^{L}_{st}(p) = -\frac{A}{F}[(1-We^{\alpha})e^{-\alpha(2N-s-t+2)} 
+(1-We^{-\alpha})e^{-\alpha(s+t-2)} ] + Ae^{-\alpha|s-t|}.
\end{equation}
Note that $\alpha$, $A$, $F$ and $W$ in the above formulas all depend on the momentum $p$ as follows:
\begin{equation}
\cosh(\alpha) = \frac{1+W^2(p) + \sum_\mu \sin^2 p_\mu}{2 |W(p)|},
\end{equation}
\begin{equation}
A = \frac{1}{2W \sinh(\alpha)},
\end{equation}
\begin{equation}
F = 1 - e^{\alpha}W(p).
\end{equation}
The form of the propagator in Eq.~(\ref{eq:dwf_prop}) is valid for positive $W$. For $0 < M_5 < 1$, $W$ is positive for any $p$. For $1 < M_5 < 2$, $W$ is negative in the small-momentum region. For that momentum region we need to adjust the formulas using the substitution
\begin{equation}
e^{\pm\alpha} \rightarrow -e^{\pm\alpha}.
\end{equation}
Using Eq.~(\ref{eq:dwf_prop}) and Eqs.~(\ref{eq:phys_quark},\ref{eq:phys_antiquark}) one can obtain the formula for the physical quark propagator
\begin{equation}
S_q(p) = \langle q(-p)\overline{q}(p) \rangle = \frac{i\gamma_\mu\sin p_\mu}{F}.
\end{equation}
For one-loop calculations we will also need propagators connecting the physical quark field and the domain-wall fermion field
\begin{equation}\label{eq:q_psibar}
\langle q(-p) \overline{\psi}_s(p)\rangle = \frac{i\gamma_\mu \sin p_\mu}{F} (e^{-\alpha(N-s)}P_R + e^{-\alpha(s-1)}P_L) - e^{-\alpha}(e^{-\alpha(N-s)}P_L + e^{-\alpha(s-1)}P_R),
\end{equation}
\begin{equation}\label{eq:psi_qbar}
\langle \psi_s(-p)\overline{q}(p)\rangle = \frac{i\gamma_\mu \sin p_\mu}{F} (e^{-\alpha(N-s)}P_R + e^{-\alpha(s-1)}P_L) - e^{-\alpha}(e^{-\alpha(N-s)}P_R + e^{-\alpha(s-1)}P_L).
\end{equation}
The relevant quark-gluon interaction vertices are the same as in the N-flavor Wilson case with $r=-1$:
\begin{equation}
V^a_{\mu}(p,p^\prime)_{st} = -igT^{a}\left(\gamma_\mu \cos\frac{(p+p^\prime)_\mu}{2} +i\sin\frac{(p+p^\prime)_\mu}{2}\right)\delta_{st},
\end{equation}
\begin{equation}
V^{ab}_{\mu\nu}(p,p^\prime)_{st} = \frac{1}{2}g^2\frac{1}{2}\{T^{a},T^{b}\} \left(i\gamma_\mu \sin\frac{(p+p^\prime)_\mu}{2} + \cos\frac{(p+p^\prime)_\mu}{2}\right)\delta_{\mu\nu}\delta_{st}.
\end{equation}
\subsection{Gauge action}
\label{subsec:gauge} 
In this paper we consider the following class of gauge actions:
\begin{equation}\label{eq:gauge_action}
S_{\rm gauge} = -\frac{2}{g^2_0} \left( (1-8c_{1}) \sum_{P} {\rm ReTr}[U_{P}] + c_{1}\sum_{R} {\rm ReTr}[U_{R}] \right) , 
\end{equation}
where $g_{0}$ denotes the bare lattice coupling, $U_{P}$ is the path-ordered product of links around the $1\times1$ plaquette $P$ and
 $U_{R}$ is the path-ordered product of links around the $1\times2$ rectangle $R$. For $c_1 = 0$, Eq.~(\ref{eq:gauge_action}) reduces to the standard Wilson plaquette action. In the case $c_1 = -1.40686$, $c_1 = -0.331$ and $c_1 = -1/12$ we have the doubly-blocked Wilson (DBW2) \cite{Takaishi:1996xj,deForcrand:1999bi}, Iwasaki \cite{Iwasaki:1983ck} and Symanzik \cite{Weisz:1983bn,Luscher:1984xn} actions respectively. 

For the Wilson plaquette action, the gluon propagator in momentum space is straightforward
\begin{equation}
D^{0}_{\nu\rho}(k) = \frac{\delta_{\nu\rho}}{4 \sum_{\mu} \sin^{2}(\frac{k_{\mu}}{2})} = \frac{\delta_{\nu\rho}}{\Delta_1} .
\end{equation}
For improved gauge action the gluon propagator is given by \cite{Weisz:1983bn} 
\begin{equation}
D^{c_1}_{\nu\rho}(k) = \frac{(1-A^{c_1}_{\nu\rho}(k))4\sin(\frac{k_{\nu}}{2})\sin(\frac{k_{\rho}}{2}) + \delta_{\nu\rho}\sum_\sigma 4\sin^2(\frac{k_{\sigma}}{2})A^{c_1}_{\nu\sigma}(k)}{\Delta^2_1} ,
\end{equation}
where $A^{c_1}_{\nu\rho}(k)$ is a function symmetric in $\nu$ and $\rho$ whose explicit dependence on $c_1$ and $k$ is given in Appendix. We assumed the Feynman gauge for both propagators.

\section{Symmetries of the action}
\label{sec:symmetries}
We will now discuss the symmetries of our action and their consequences for heavy-light operators.
Similar discussions for different choices of action were presented in Refs.~\cite{Becirevic:2003hd,Yamada:2004ri}. We will use the following notation to denote lattice heavy-light bilinears
\begin{equation}
O^{\rm lat}_\Gamma = \bar{h}\Gamma q = \left\{ \begin{array}{ll}
 S  & (\Gamma = 1) \\
 P & (\Gamma = \gamma_5) \\
 V_\mu & (\Gamma = \gamma_\mu) \\
 A_\mu & (\Gamma = \gamma_\mu\gamma_5) \\
 T_{\mu\nu} & (\Gamma = \sigma_{\mu\nu} )
\end{array}
\right. .
\end{equation}

\subsection{Chiral symmetry}
The five-dimensional domain-wall fermion field $\psi_s(x)$ can be subjected to the following chiral rotations:
\begin{equation}\label{eq:chi_rotL}
\psi_{s}(x) \to e^{i\alpha^{a}_Lt^{a}}\psi_{s}(x),  \quad  1 \leqslant s \leqslant N/2 ,
\end{equation}
\begin{equation}\label{eq:chi_rotR}
\psi_{s}(x) \to e^{i\alpha^{a}_Rt^{a}}\psi_{s}(x),  \quad  N/2+1 \leqslant s \leqslant N ,
\end{equation}
where $\{t_{a}\}$ are the generators of the ${\rm SU}(N_{f})$ flavor group and $N$ is the extent of the fifth dimension. Together with the definition of the physical quark field $q$ in Eq.~(\ref{eq:phys_quark}), the transformations in Eqs.~(\ref{eq:chi_rotL},\ref{eq:chi_rotR}) act as standard chiral symmetry transformations on $q$. 
For finite $N$ this symmetry is not exact. However, as was shown in Refs.~\cite{Blum:2000kn,Capitani:2006kw} the degree of the symmetry breaking decreases rapidly with increasing $N$.   
Since we take $N \to \infty$ limit in this calculation, we can consider the symmetry breaking to be negligible. Thus, under chiral symmetry transformations we have $S \to iP$, $V_\mu \to iA_\mu$ and vice versa. This implies that at the level of the static theory-to-lattice matching we should expect $Z_S = Z_P$ and $Z_V=Z_A$, where $Z$'s are defined in Eq.~(\ref{eq:bilin_match}).  

\subsection{Heavy quark symmetries}
First, we note that since the static-quark field $h$ satisfies the field equation $\gamma_0 h = h$ we have $V_0 = S$ and $A_0 = P$ as well as $T_{0j} = V_j$ and $T_{ij} = \epsilon_{ijk}A_k$.
Second, the action in Eq.~(\ref{eq:hqa}) as well as the continuum static-effective-field-theory action are invariant under the following {\rm SU}(2) transformation
\begin{equation}
h \to e^{-i\phi_{j}\epsilon_{jkl}\sigma_{kl}} h,
\end{equation}
where $\phi_{j}$ is a parameter and $\sigma_{kl} = \frac{i}{2}[\gamma_{k}, \gamma_{l}]$. 
This is the heavy-quark spin symmetry \cite{Isgur:1989vq,Isgur:1989ed}. As its consequence we have the following transformations
$S \to iA_j$, $P \to iV_j$ and vice versa. This implies that at the level of the static theory-to-lattice matching we should expect $Z_S = Z_A$ and $Z_P=Z_V$. 

\section{Matching full-theory and static-theory operators}
\label{sec:match_full_effective_op}
In this section we will briefly review the relation of full QCD operators renormalized in ${ \rm \overline{MS}}$(NDR) scheme to the corresponding static theory operators. 
\subsection{Heavy-light bilinears}
The QCD bilinear $O_\Gamma = \bar{b}\Gamma q$, renormalized in ${ \rm \overline{MS}}$(NDR) at the scale $\mu_b$, can be related to the corresponding static theory bilinear $\tilde{O}_\Gamma = \bar{h}\Gamma q$, renormalized at the scale $\mu$, as follows \cite{Shifman:1986sm,Politzer:1988wp}:
\begin{equation}
O_\Gamma(\mu_b) = C_{O_\Gamma}(\mu_b, \mu)\tilde{O}_\Gamma(\mu) + {\mathcal O}(\Lambda_{QCD}/\mu_b), 
\end{equation}
where $C_{O_\Gamma}(\mu_b, \mu)$ is a perturbative coefficient encoding the physics between the scales $\mu_b$ and $\mu$. In lattice calculations $\mu$ naturally assumes the value of the inverse lattice spacing and $\mu_b$ is chosen to match the $b$ quark mass. The coefficients $C_{O_\Gamma}$ for each $O_\Gamma$ were calculated perturbatively at one-loop level in Ref.~\cite{Eichten:1989zv} and at two-loop level in Ref.~\cite{Broadhurst:1994se}. The coefficients $C_{O_\Gamma}$ are different for different bilinears.      

\subsection{Four-fermion operator}
The QCD operator relevant for $B^{0} - \bar{B}^{0}$ mixing, here renormalized according to the $\overline{\rm MS}$(NDR) scheme, is 
\begin{equation}\label{eq:msndr_v-a} O_{(V-A)(V-A)} = \left[\bar{b}\gamma^{\mu}(1-\gamma_{5})q\right]\left[\bar{b}\gamma_{\mu}(1-\gamma_{5})q\right] . \end{equation}
We need to consider only the parity conserving part of it:
\begin{equation}\label{eq:msndr_vvaa}
O_{VV+AA} = \left(\bar{b}\gamma^{\mu}q\right)\left(\bar{b}\gamma_{\mu}q\right) + \left(\bar{b}\gamma^{\mu}\gamma_{5}q\right)\left(\bar{b}\gamma_{\mu}\gamma_{5}q\right) . 
\end{equation} 
The relation to the static theory operators is as follows:
\begin{equation}\label{eq:full_vv+aa_matching}
 O_{VV+AA}(\mu_{b}) = Z_{1}(\mu_{b},\mu) \tilde{O}_{VV+AA}(\mu) +  Z_{2}(\mu_{b},\mu)\tilde{O}_{SS+PP}(\mu) +  {\mathcal O}(\Lambda_{QCD}/\mu_{b}),
\end{equation}
where
\begin{equation}
\tilde{O}_{VV+AA} = 2\left(\bar{h}^{(+)}\gamma^{\mu}q\right)\left(\bar{h}^{(-)}\gamma_{\mu}q\right) + 2\left(\bar{h}^{(+)}\gamma^{\mu}\gamma_{5}q\right)\left(\bar{h}^{(-)}\gamma_{\mu}\gamma_{5}q\right),
\end{equation}
\begin{equation}
\tilde{O}_{SS+PP} = 2\left(\bar{h}^{(+)}q \right)\left(\bar{h}^{(-)}q\right) + 2\left(\bar{h}^{(+)}\gamma_{5}q\right)\left(\bar{h}^{(-)}\gamma_{5}q\right),
\end{equation}
with
\begin{equation}
h^{(\pm)}(x) = e^{\pm imv\cdot x}\frac{1 \pm \slash{v}}{2}b(x) .
\end{equation}
The coefficients $Z_{1,2}$ were calculated at one-loop in perturbation theory in Refs.~\cite{Gimenez:1992is,Ciuchini:1996sr,Buchalla:1996ys}. We note that $Z_2$ is an order $g^2$ coefficient.

\section{Matching static-theory and lattice heavy-light bilinears}
\label{sec:match_effective_lattice_op}
In this section we will match the $\overline{\rm MS}$(NDR) renormalized static theory heavy-light bilinears introduced in Sec.~\ref{sec:match_full_effective_op} to their lattice counterparts. We will use the lattice Feynman rules from Sec.~\ref{sec:action}.

The Feynman graphs representing one-loop corrections to the tree-level heavy-light bilinears are depicted in Fig.~\ref{fig:current_corrections}. 
{\unitlength=1.2mm
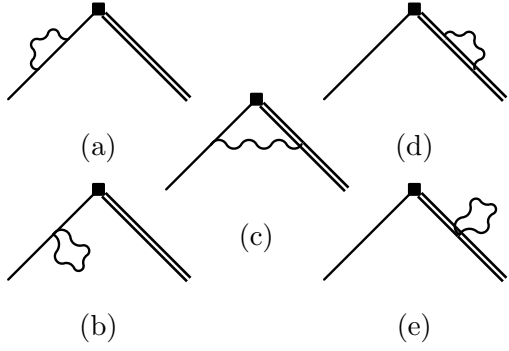
\begin{figure}[h]
\centering
\begin{fmffile}{vertex}
\begin{fmfgraph*}(50,50)
\begin{fmfsubgraph}(0,.4h)(.4w,.4h)
\begin{fmfgroup}
\fmfset{arrow_len}{cm}\fmfset{arrow_ang}{5}
\fmfright{i_b}
\fmftop{x_b}
\fmfleft{o_b}
\fmfiv{l=(a),l.a=-90,l.d=.075w}{c}
\fmf{double}{i_b,x_b}
\fmf{plain}{x_b,v_b}
\fmf{plain}{v_b,v1_b}
\fmf{wiggly,right,tension=0}{v_b,v1_b}
\fmf{plain}{v1_b,o_b}
\fmfv{decor.shape=square,decor.filled=1,decor.size=2thick}{x_b}
\end{fmfgroup}
\end{fmfsubgraph}
\begin{fmfsubgraph}(0,0)(.4w,.4h)
\begin{fmfgroup}
\fmfset{arrow_len}{cm}\fmfset{arrow_ang}{5}
\fmfright{i_c}
\fmftop{x_c}
\fmfleft{o_c}
\fmfiv{l=(b),l.a=-90,l.d=.075w}{c}
\fmf{plain}{x_c,v_c}
\fmf{wiggly,right}{v_c,v_c}
\fmf{plain}{v_c,o_c}
\fmf{double}{i_c,x_c}
\fmfv{decor.shape=square,decor.filled=1,decor.size=2thick}{x_c}
\end{fmfgroup}
\end{fmfsubgraph}
\begin{fmfsubgraph}(0.35w,.2h)(.4w,.4h)
\begin{fmfgroup}
\fmfright{i_a}
\fmftop{x_a}
\fmfleft{o_a}
\fmfiv{l=(c),l.a=-90,l.d=.075w}{c}
\fmf{double}{i_a,v_a}
\fmf{double}{v_a,x_a}
\fmf{plain}{x_a,v1_a}
\fmf{plain}{v1_a,o_a}
\fmf{wiggly,tension=0}{v_a,v1_a}
\fmfv{decor.shape=square,decor.filled=1,decor.size=2thick}{x_a}
\end{fmfgroup}
\end{fmfsubgraph}
\begin{fmfsubgraph}(.7w,.4h)(.4w,.4h)
\fmfright{i_d}
\fmftop{x_d}
\fmfleft{o_d}
\fmfiv{l=(d),l.a=-90,l.d=.075w}{c}
\fmf{double}{i_d,v_d}
\fmf{double}{v_d,v1_d}
\fmf{wiggly,right,tension=0}{v_d,v1_d}
\fmf{double}{v1_d,x_d}
\fmf{plain}{x_d,o_d}
\fmfv{decor.shape=square,decor.filled=1,decor.size=2thick}{x_d}
\end{fmfsubgraph}
\begin{fmfsubgraph}(.7w,0)(.4w,.4h)
\fmfright{i_e}
\fmftop{x_e}
\fmfleft{o_e}
\fmfiv{l=(e),l.a=-90,l.d=.075w}{c}
\fmf{double}{i_e,v_e}
\fmf{double}{v_e,x_e}
\fmfset{curly_len}{2mm}
\fmf{wiggly,right}{v_e,v_e}
\fmf{plain}{x_e,o_e}
\fmfv{decor.shape=square,decor.filled=1,decor.size=2thick}{x_e}
\end{fmfsubgraph}
\end{fmfgraph*}
\end{fmffile}
\caption{\label{fig:current_corrections} One-loop corrections to heavy-light bilinears.}
\end{figure}
}
The light-quark wave function renormalization factor coming from the diagrams (a) and (b) has been calculated in Refs.~\cite{Aoki:1998vv,Aoki:2002iq}. The vertex correction (c) has not been calculated before for the case of DWF-static quarks. We will present this calculation below. The heavy-quark wave function renormalization factor coming from the diagrams (d) and (e) has been calculated for the Wilson gauge action in Ref.~\cite{Eichten:1989kb}. We will extend the result of Ref.~\cite{Eichten:1989kb} to include the case of improved gauge actions.

\subsection{Vertex correction}
At the tree level, the Green's function $\langle (\overline{h}(x)\Gamma q(x))h(y)\overline{q}(z)\rangle$ for small external momenta $p$ is given by 
\begin{equation}
\langle (\overline{h}\Gamma q)h\overline{q}\rangle_{\rm tree} = \frac{1}{ip_0} \Gamma \frac{(1-w^2_0)}{i \slash{p}} ,
\end{equation}
where
\begin{equation}
w_0 = W(0) = 1 - M_5. 
\end{equation}
We need to calculate the one-loop correction to this Green's function coming from the graph depicted in Fig.~\ref{fig:vertex_graph}.
{
\unitlength=1.2mm
\begin{figure}[h]
\centering
\begin{fmffile}{vertex_details}
\begin{fmfgraph*}(50,50)
\begin{fmfsubgraph}(0w,-0.3h)(1.0w,1.0h)
\fmfleft{i_a}
\fmftop{x_a}
\fmfright{o_a}
\fmflabel{$\bar q$}{o_a}
\fmflabel{$\bar \psi_s \psi_s$}{v1_a}
\fmflabel{$\bar h \Gamma q$}{x_a}
\fmflabel{$h$}{i_a}
\fmf{heavy}{x_a,v_a}
\fmf{heavy}{v_a,i_a}
\fmf{fermion}{o_a,v1_a}
\fmf{fermion}{v1_a,x_a}
\fmf{wiggly,tension=0}{v_a,v1_a}
\fmfv{decor.shape=square,decor.filled=1,decor.size=4thick}{x_a}
\fmfdot{v_a}
\fmfdot{v1_a}
\end{fmfsubgraph}
\end{fmfgraph*}
\end{fmffile}
\caption{\label{fig:vertex_graph} Heavy-light vertex correction}
\end{figure}
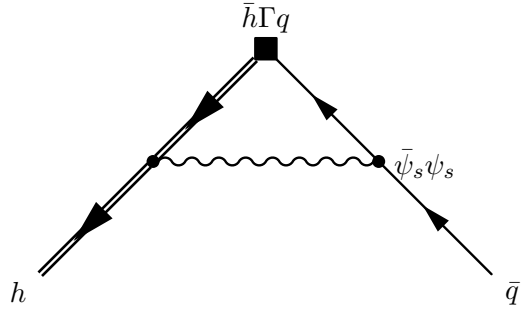
}
As can be seen from Eq.~(\ref{eq:psi_qbar}), for small external momenta the external line of the light quark propagator is 
\begin{equation}
\langle \psi_s(-p\approx 0)\overline{q}(p\approx 0)\rangle = (w_0^{N-s}P_L + w_0^{s-1}P_R) \frac{(1-w^2_0)}{i \slash{p}}.
\end{equation}
This implies that the one-loop vertex-corrected Green's function will take the form
\begin{equation}
\langle (\overline{h}\Gamma q)h\overline{q}\rangle = \frac{1}{ip_0} (\Gamma +\delta\Gamma) \frac{(1-w^2_0)}{i \slash{p}} ,
\end{equation} 
where 
\begin{equation}\label{eq:dg_basic}
\delta\Gamma = \int_{k} \sum_{\nu,\rho} H(k) \tilde{V}^a_\nu(k) D^{c_1}_{\nu\rho}(k) \Gamma \left\{\sum^N_{s=1} \langle q(-k)\overline{\psi_s}(k)\rangle  V^a_{\rho}(k) (w_0^{N-s}P_L + w_0^{s-1}P_R)\right\} , 
\end{equation}
and
\begin{equation}
\int_k \equiv \int_{-\pi}^{\pi} \frac{d^4k}{(2\pi)^4} .
\end{equation}
First, we focus on the factor in curly brackets in the integrand of Eq.~(\ref{eq:dg_basic}). Since the light quark-gluon vertex $V^a_{\rho}$ has $\gamma_\rho$ and Wilson parts, it is convenient to consider them separately. We denote the part coming from the $\gamma_\rho$ term as $S_\chi$ and the part coming from the Wilson term as $S_w$. Then   
\begin{equation}
S_\chi(k) = \sum^N_{s=1} \langle q(-k)\overline{\psi_s}(k)\rangle (w_0^{N-s}P_R + w_0^{s-1}P_L) ,
\end{equation}
\begin{equation}
S_w(k) = \sum^N_{s=1} \langle q(-k)\overline{\psi_s}(k)\rangle (w_0^{N-s}P_L + w_0^{s-1}P_R) .
\end{equation}
Using Eqs.~(\ref{eq:q_psibar},\ref{eq:psi_qbar}) and performing the sum over $s$ we find
\begin{equation}
S_\chi(k) =  \frac{1}{w_0 - e^{\alpha}}\frac{i \sum_\mu \gamma_\mu\sin k_\mu}{W(k) - e^{-\alpha}} ,
\end{equation}
\begin{equation}
S_w(k) = \frac{1}{w_0 - e^{\alpha}} .
\end{equation}
Combining this result with the Feynman rules presented in Sec.~\ref{sec:action},
 the one-loop correction to the vertex in Fig.~\ref{fig:vertex_graph} can be written as 
\begin{equation}\label{eq:dwf_vertex}
\delta\Gamma = -g^2 C_F \int_{k} \sum_\rho \frac{e^{-i\frac{k_0}{2}}}{1- e^{-ik_0} + \epsilon} D^{c_1}_{0\rho}(k) \Gamma \left\{S_\chi(k) \gamma_\rho \cos(\frac{k_\rho}{2}) + S_w(k) i \sin(\frac{k_\rho}{2})\right\},
\end{equation}
where the second Casimir $C_F = (N^2_c-1)/2N_c$.
Performing the sum over $\rho$ and dropping the terms odd in $k$ we can write it as follows: 
\begin{equation}
\delta\Gamma = \Gamma \delta V_\Gamma = -\frac{g^2}{16\pi^2}C_F\Gamma (I_\chi + I_w) ,
\end{equation}
where 
\begin{equation}\label{eq:I_chi_def}
I_\chi = 16\pi^2 \int_{k} \frac{\sin^2(k_0) + 4\sum_j A^{c_1}_{0j}\sin^2(\frac{k_j}{2})\cos^2(\frac{k_0}{2}) + \sum_j (1 - A^{c_1}_{0j})\sin^2(k_j)}{(w_0 - e^{\alpha})(W(k) - e^{-\alpha})\Delta^2_1} ,
\end{equation}
and
\begin{equation}
I_w = 16\pi^2 \int_{k} \frac{1}{2 (w_0-e^{\alpha}) \Delta_1}.
\end{equation}
The integral $I_w$ is finite and can be evaluated numerically. The integral $I_\chi$ is infrared divergent. We can write it as follows: 
\begin{equation}
I_\chi = I_\chi^{c_1} + I_\chi^w ,
\end{equation}
where the first term is the finite contribution coming from the difference between the improved and the Wilson gauge action 
\begin{equation}\label{eq:I_chi_def_diff}
I_\chi^{c_1} = 16\pi^2 \int_{k} \left\{G(c_1,k)-G(0,k)\right\},
\end{equation}
with $G(c_1,k)$ representing the integrand in Eq.~(\ref{eq:I_chi_def}).
The second term
\begin{equation}
I_\chi^w = 16\pi^2 \int_k G(0,k) =  16\pi^2 \int_k \frac{\cos^2(\frac{k_0}{2})}{(w_0 - e^{\alpha})(W(k) - e^{-\alpha})\Delta_1},
\end{equation}
is just the Wilson gauge action case added back in. Now, the integral $I_\chi^w$ is infrared divergent. We will regularize it by including a gluon mass $\lambda$ and extract the finite part using the method described in Ref.~\cite{Bernard:1987rw}. We have
\begin{equation}
I_\chi^w = {\mathcal I}_\chi^w - \hat{\mathcal I}_\chi^w - 16\pi^2\int_k \frac{\theta(1- k^2)}{k^2(k^2+\lambda^2)} , 
\end{equation}
where 
\begin{equation}
{\mathcal I}^w_\chi =  16\pi^2 \int_k \left( \frac{1}{(w_0 - e^{\alpha})(W(k) - e^{-\alpha})\Delta_1} + \frac{\theta(1- k^2)}{k^4} \right ),
\end{equation}
and 
\begin{equation}
\hat{\mathcal I}_\chi^w = 16\pi^2 \int_k \frac{1}{16(w_0 - e^{\alpha})(W(k) - e^{-\alpha})};
\end{equation}
both are finite and can be calculated numerically.

Then the one-loop correction to the vertex is given by
\begin{equation}\label{eq:vertex_result}
\delta V_\Gamma = \frac{g^2}{16\pi^2}C_F[-\ln(\lambda^2 a^2) - I^{c_1}_\chi - {\mathcal I}^w_\chi + \hat{\mathcal I}^w_\chi - I_w ] =  \frac{g^2}{16\pi^2}C_F[-\ln(\lambda^2 a^2) + d].
\end{equation}
Note that neither the constant $d$ nor the coefficient of the logarithm depend on the matrix $\Gamma$ at the vertex. The numerical results for $I_w$, ${\mathcal I}_\chi^w$ and $\hat{\mathcal I}_\chi^w$ are tabulated in Tab.~\ref{tab:dwf_tad}. The values of $I^{c_1}_\chi$ for different gauge actions can be found in Tab.~\ref{tab:dwf_i-w}. The value of the constant $d$ is given in Tab.~\ref{tab:vertex}. These and all other numerical integrals in this paper were calculated using the Monte Carlo integration routine VEGAS\cite{Lepage:1977sw}, with 1000000 sample points.

\subsection{Heavy quark corrections}
Now, let's focus on the heavy quark propagator corrections. The tadpole graph in Fig.~\ref{fig:current_corrections}e is given by the following expression 
\begin{equation}
\Sigma^{\rm tad}(p_0) = -\frac{1}{2}g^2C_F e^{-ip_0} \int_k D^{c_1}_{00}(k) =  -\frac{1}{2}g^2C_Fe^{-ip_0} T_{c_1},
\end{equation}
where $T_{c_1}$ is the tadpole integral whose numerical values are 
\begin{equation}\label{eq:4d_tadpole}
T_{c_1} = \left\{
\begin{array}{ll}
 0.154933(3) & {\rm Wilson, c_1 = 0} \\
 0.094759(3) & {\rm Iwasaki, c_1 = -0.331} \\
 0.062426(2) & {\rm DBW2, c_1 = -1.40686} \\
 0.128291(3) & {\rm Symanzik, c_1 = -1/12 }
\end{array}
\right..
\end{equation}
The formula for the rising sun diagram in Fig.~\ref{fig:current_corrections}d is given by
\begin{equation}\label{eq:sigma_rs}
\Sigma^{\rm rs}(p_0) = -g^{2}C_F \int_k D^{c_1}_{00}(k-p)\frac{e^{-i(p_0+k_0)}}{1- e^{-ik_0} + \epsilon}.
\end{equation}
The heavy quark wave function renormalization is obtained via
\begin{equation}
Z^h_2 - 1 = -i\frac{\partial(\Sigma^{\rm tad}+\Sigma^{\rm rs})}{\partial p_0}|_{p_0=0}.
\end{equation}
The calculation of the tadpole contribution to $Z^h_2$ is trivial. In order to calculate the contribution from $\Sigma^{\rm rs}$ we will use a method similar to the one developed in Ref.~\cite{Eichten:1989kb}. We write
\begin{equation}\label{eq:sigma_deriv_split}
-i\frac{\partial\Sigma^{\rm rs}}{\partial p_0}|_{p_0=0} = \Sigma_{\rm exp} + \Sigma_D,
\end{equation}
where $\Sigma_{\rm exp}$ and $\Sigma_D$ come from differentiating $\exp[-i(p_0 + k_0)]$ and $D^{c_1}_{00}(k-p)$ in Eq.~(\ref{eq:sigma_rs}) respectively. We get
\begin{equation}\label{eq:stat_vrtx_part}
\Sigma_{\rm exp} = g^{2}C_F \int_k D^{c_1}_{00}(k) \frac{e^{-ik_0}}{1 - e^{-ik_0} + \epsilon}.
\end{equation}
Note that 
\begin{equation}\label{eq:cot}
\frac{e^{-ik_0}}{1 - e^{-ik_0}} = \frac{1}{2i}\cot(\frac{k_0}{2}) -\frac{1}{2}.  
\end{equation}
Then Eq.~(\ref{eq:stat_vrtx_part}) can be simplified to  
\begin{equation}
\Sigma_{\rm exp} = \frac{1}{2}g^2C_F T^{(3)}_{c_1} - \frac{1}{2}g^2C_F T_{c_1},
\end{equation}
where
\begin{equation}
T^{(3)}_{c_1} = \int_{-\pi}^{\pi} \frac{d^3k}{(2\pi)^3} D^{c_1}_{00}(0,\vec{k}).
\end{equation}
The numerical results for $T^{(3)}_{c_1}$ are
\begin{equation}
T^{(3)}_{c_1} = \left\{
\begin{array}{ll}
 0.252721(6) & {\rm Wilson, c_1 = 0} \\
 0.164361(5) & {\rm Iwasaki, c_1 = -0.331} \\
 0.093831(4) & {\rm DBW2, c_1 = -1.40686} \\
 0.219018(6) & {\rm Symanzik, c_1 = -1/12 }
\end{array}
\right..
\end{equation}

For $\Sigma_D$ in Eq.~(\ref{eq:sigma_deriv_split}) we obtain 
\begin{equation}\label{eq:stat_glue_part}
\Sigma_D = -ig^{2}C_F \int_k D^{c_1\prime}_{00}(k) 4\sin(\frac{k_0}{2})\cos(\frac{k_0}{2}) \frac{e^{-ik_0}}{1- e^{-ik_0} + \epsilon},
\end{equation}
where
\begin{equation}
D^{c_1\prime}_{00}(k) = \frac{\partial D^{c_1}_{00}(\widehat{k}_0^2,\vec{k})}{\partial \widehat{k}_0^2}, 
\end{equation}
with $\widehat{k}_0 = 2\sin(k_0/2)$. We calculated $D^{c_1\prime}_{00}(k)$ numerically and symbolically using Mathematica. Using Eq.~(\ref{eq:cot}) and dropping the term odd in $k_0$ we can reduce Eq.~(\ref{eq:stat_glue_part}) to   
\begin{equation}\label{eq:rainbow_2nd_part}
\Sigma_D = -2 g^{2}C_F\int_k D^{c_1\prime}_{00}(k) \cos^2(\frac{k_0}{2}). 
\end{equation}
Again, we calculate the difference between the improved and the Wilson gauge action numerically: 
\begin{equation}
I^{c_1}_{h} =  16\pi^2 \int_k \left(D^{c_1\prime}_{00}(k)  + \frac{1}{\Delta^2_1} \right )\cos^2(\frac{k_0}{2}).
\end{equation}
The results are
\begin{equation}
I^{c_1}_h = \left\{
\begin{array}{ll}
 1.310153(55) & {\rm Iwasaki, c_1 = -0.331} \\
 3.510284(162) & {\rm DBW2, c_1 = -1.40686} \\
 0.297720(15) & {\rm Symanzik, c_1 = -1/12 }
\end{array}
\right..
\end{equation}
The contribution of the Wilson part to be added back in is
\begin{equation}
16 \pi^2 \int_k \frac{\cos^2(\frac{k_0}{2})}{\Delta^2_1} = 16 \pi^2 \int_k \frac{1}{\Delta^2_1} - \pi^2\int_k \frac{1}{\Delta_1} =  16 \pi^2 \int_k \frac{1}{\Delta^2_1} - \pi^2 T_0,
\end{equation}
where $T_0$ is the Wilson tadpole integral calculated in Eq.~(\ref{eq:4d_tadpole}).
Once again, we subtract numerically the infrared divergent part using the method of Ref.~\cite{Bernard:1987rw}:\begin{equation}
\Theta = 16\pi^2 \int_k \left( \frac{1}{\Delta^2_1} - \frac{\theta(1-k^2)}{k^4} \right) = 4.791861(251).
\end{equation}
Finally, putting together all one-loop contributions to $Z_2^h$ we have 
\begin{equation}\label{eq:Zh_final}
Z_2^h  = 1+ 2\frac{g^2}{16\pi^2}C_F[-\ln(\lambda^2 a^2) -1 -I^{c_1}_h + \Theta + 4\pi^2 T^{(3)}_{c_1} - \pi^2T_0], 
\end{equation}
or
\begin{equation}
Z_2^h = 1 + \frac{g^2}{16\pi^2}C_F[-2\ln(\lambda^2 a^2) + e],
\end{equation}
where
\begin{equation}\label{eq:e_results}
e = \left\{
\begin{array}{ll}
 24.480 & {\rm Wilson, c_1 = 0} \\
 14.883 & {\rm Iwasaki, c_1 = -0.331} \\
 4.914 & {\rm DBW2, c_1 = -1.40686} \\
 21.223 & {\rm Symanzik, c_1 = -1/12 }
\end{array}
\right.,
\end{equation}
with a maximum error of $\pm 1$ in the last digit.
Our result for the Wilson case agrees with the result of Ref.~\cite{Eichten:1989kb}. Note that the correction to $Z_2^h$ is significantly reduced in the case of Iwasaki and DBW2 action.  

Now, let us discuss the radiative correction to the mass 
\begin{equation}
\delta M = - \Sigma^{\rm tad}(p_0=0) - \Sigma^{\rm rs}(p_0=0).
\end{equation}
We obtain 
\begin{equation}
\delta M = \frac{g^2}{16 \pi^2}C_F 8\pi^2T^{(3)}_{c_1} = \frac{g^2}{16 \pi^2}C_F \times
 \left\{
\begin{array}{ll}
 19.954 & {\rm Wilson, c_1 = 0} \\
 12.977 & {\rm Iwasaki, c_1 = -0.331} \\
 7.409 & {\rm DBW2, c_1 = -1.40686} \\
 17.293 & {\rm Symanzik, c_1 = -1/12 }
\end{array}
\right.,
\end{equation}
which agrees with the Wilson case result in Refs.~\cite{Eichten:1989kb,Boucaud:1992nf,Aglietti:1993hf}. 

\subsection{Relation to $\overline{\bf MS}$ scheme}
Now, we can calculate $Z_\Gamma$ relating the continuum static theory bilinear $\tilde{O}_\Gamma$, renormalized in ${\rm \overline{MS}(NDR)}$ , to the lattice bilinear $O^{\rm lat}_\Gamma$ as follows:
\begin{equation}\label{eq:bilin_match}
\tilde{O}_\Gamma(\mu) = (1-w^2_0)^{-1/2}Z^{-1/2}_w Z_\Gamma(\mu,a) O^{\rm lat}_\Gamma(a),
\end{equation}
where $(1-w^2_0)^{-1/2}Z^{-1/2}_w$ is a DWF-specific factor, whose origin and numerical values are discussed in Refs.~\cite{Aoki:1998vv,Aoki:2002iq} and 

\begin{multline}\label{eq:zh_prelim}
Z_\Gamma(\mu,a) =  1 + \frac{g^2}{16\pi^2}C_F(-\ln\frac{\lambda^2}{\mu^2} + D + \frac{1}{2}(-2\ln\frac{\lambda^2}{\mu^2} + E) + \frac{1}{2}(\ln\frac{\lambda^2}{\mu^2} + F) ) \\ -\frac{g^2}{16\pi^2}C_F(-\ln\lambda^2a^2 + d + \frac{1}{2}(-2\ln\lambda^2a^2 + e) + \frac{1}{2}(\ln\lambda^2a^2 + f)).  
\end{multline}
The continuum static theory constants $D=1$, $E=0$ and $F=1/2$ were calculated in Ref.~\cite{Eichten:1989zv}. The values for $d$ and $e$ can be found in Tab.~\ref{tab:vertex} and Eq.~(\ref{eq:e_results}). The lattice light quark renormalization factor $f = 1/2-z_2$, where $z_2$ was calculated in Refs.~\cite{Aoki:1998vv,Aoki:2002iq}. Eliminating $\lambda$ from Eq.~(\ref{eq:zh_prelim}) we obtain    
\begin{equation}\label{eq:Za_final}
Z_\Gamma(\mu, a) = 1 + \frac{g^2}{16\pi^2}C_F(\frac{3}{2} \ln \mu^2a^2 + \frac{5}{4} - d - \frac{e}{2} - \frac{f}{2}).
\end{equation}  
Note that the result for $Z_\Gamma$ does not depend on $\Gamma$ since both $D$ and $d$ are $\Gamma$ independent. This is consistent with the observation in Sec.~\ref{sec:symmetries} that $Z_\Gamma$ should be the same for all bilinears because of the symmetries of the action.

\section{Matching static-theory and lattice four-fermion operators}
\label{sec:match_effective_lattice_4ferm_op}
In Sec.~\ref{sec:match_full_effective_op} we discussed the relationship between the full theory operator $O_{VV+AA}$ and static theory operators $\tilde{O}_{VV+AA}$ and $\tilde{O}_{SS+PP}$. In this section we will match these static theory operators renormalized in $\overline{\rm MS}$(NDR) scheme to lattice operators. We will also discuss the renormalization of $B_B$.
\subsection{Operators matching}
 First, we note that the coefficient $Z_2$ in Eq.~(\ref{eq:full_vv+aa_matching}) is already of the order $g^2$. Thus, for the purposes of matching at the leading order in $g^2$ it will suffice to match $\tilde{O}_{SS+PP}$ and $O^{\rm lat}_{SS+PP}$ at tree level.
\begin{equation}
\tilde{O}_{SS+PP}(\mu) =  Z_S O^{\rm lat}_{SS+PP}(a),
\end{equation}
with
\begin{equation}
Z_S = (1-w^2_0)^{-1}Z^{-1}_w,
\end{equation}
being the DWF-specific factor present even at tree level.
The matching of $\tilde{O}_{VV+AA}$  to the lattice has been worked out for the Wilson case in Refs.~\cite{Flynn:1990qz,Gimenez:1998mw}. In our case, the mixing is simplified, since the chiral symmetry prohibits mixing with operators of different chirality \cite{Capitani:2000da,Becirevic:2003hd,Becirevic:2005sx}. With the exception of the diagrams depicted in Fig.~\ref{fig:4ferm_corrections}, the four-fermions diagrams can be reduced to the diagrams already discussed in Sec.~\ref{sec:match_effective_lattice_op}. Thus, we have
\begin{equation}
\tilde{O}_{VV+AA}(\mu) = Z_L O^{\rm lat}_{VV+AA}(a), 
\end{equation} 
with
\begin{equation}
Z_L = (1-w^2_0)^{-1}Z^{-1}_w \left(1 + \frac{g^2}{16\pi^2}\left(4 \ln\mu^2a^2 + D_L \right)\right),
\end{equation}
and
\begin{equation}\label{eq:d_l}
D_L = \frac{7}{3} -\frac{1}{3}c -\frac{10}{3}d -\frac{4}{3}e -\frac{4}{3}f -\frac{1}{3}v.
\end{equation}
The constants $d$, $e$ and $f$ were discussed in Sec.~\ref{sec:match_effective_lattice_op}. The constant $v$ is the light-light correction depicted in Fig.~\ref{fig:4ferm_corrections}a, which was calculated in Refs.~\cite{Aoki:1998vv,Aoki:1999ky,Aoki:2002iq}. In the notation of Ref.~\cite{Aoki:1998vv} $v = -V_{S,P}$. The constant $c$ is the heavy-heavy correction depicted in Fig.~\ref{fig:4ferm_corrections}b, which we calculate below.
{
\unitlength=1.2mm
\begin{figure}[h]
\centering
\begin{fmffile}{b_mixing}
\begin{fmfgraph*}(50,50)
\begin{fmfsubgraph}(0,.5h)(.4w,.4h)
\begin{fmfgroup}
\fmfset{arrow_len}{cm}\fmfset{arrow_ang}{5}
\fmfright{i_b}
\fmfbottom{x_b}
\fmfleft{o_b}
\fmf{double}{i_b,x_b}
\fmf{plain}{x_b,v_b}
\fmf{plain}{v_b,v1_b}
\fmf{wiggly,right,tension=0}{v1_b,v_c}
\fmf{plain}{v1_b,o_b}
\fmfv{decor.shape=square,decor.filled=1,decor.size=3thick}{x_b}
\end{fmfgroup}
\end{fmfsubgraph}
\begin{fmfsubgraph}(0,0)(.4w,.4h)
\begin{fmfgroup}
\fmfset{arrow_len}{cm}\fmfset{arrow_ang}{5}
\fmfright{i_c}
\fmftop{x_c}
\fmfleft{o_c}
\fmfiv{l=(a),l.a=-90,l.d=.075w}{c}
\fmf{plain}{x_c,v_c}
\fmf{plain}{v_c,o_c}
\fmf{double}{i_c,x_c}
\fmfv{decor.shape=square,decor.filled=1,decor.size=3thick}{x_c}
\end{fmfgroup}
\end{fmfsubgraph}
\begin{fmfsubgraph}(.5w,.5h)(.4w,.4h)
\fmfright{i_d}
\fmfbottom{x_d}
\fmfleft{o_d}
\fmf{double}{i_d,v_d}
\fmf{double}{v_d,x_d}
\fmf{plain}{x_d,o_d}
\fmfv{decor.shape=square,decor.filled=1,decor.size=3thick}{x_d}
\end{fmfsubgraph}
\begin{fmfsubgraph}(.5w,0)(.4w,.4h)
\fmfright{i_e}
\fmftop{x_e}
\fmfleft{o_e}
\fmfiv{l=(b),l.a=-90,l.d=.075w}{c}
\fmf{double}{i_e,v_e}
\fmf{wiggly,left,tension=0}{v_d,v_e}
\fmf{double}{v_e,x_e}
\fmf{plain}{x_e,o_e}
\fmfv{decor.shape=square,decor.filled=1,decor.size=3thick}{x_e}
\end{fmfsubgraph}
\end{fmfgraph*}
\end{fmffile}
\caption{\label{fig:4ferm_corrections} Four-fermion specific one-loop corrections.}
\end{figure}
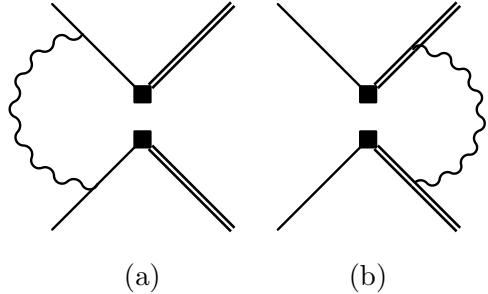
}

Omitting the trivial $\Gamma$ structure, the contribution from the graph in Fig.~\ref{fig:4ferm_corrections}b is given by the following expression:
\begin{equation}\label{eq:hh_corr_def}
\delta V = g^2C_F \int_k D^{c_1}_{00}(k)\frac{e^{-ik_0}}{(1-e^{-ik_0}+\epsilon)^2}.
\end{equation}
Noting that
\begin{equation} \frac{e^{-ik_0}}{(1-e^{-ik_0})^2} = -\frac{1}{4\sin^2(k_0/2)},
\end{equation}
we integrate Eq.~(\ref{eq:hh_corr_def}) by parts and arrive at
\begin{equation}\label{eq:hh_final_form}
\delta V = -2g^2C_F \int_k D^{c_1\prime}_{00}(k)\cos^2(\frac{k_0}{2}),
\end{equation} which is identical to Eq.~(\ref{eq:rainbow_2nd_part}) in Sec.~\ref{sec:match_effective_lattice_op}. Using our results from Sec.~\ref{sec:match_effective_lattice_op} we can immediately write  
\begin{equation}
\delta V = 2\frac{g^2}{16\pi^2}C_F[-\ln\lambda^2a^2 - 1 - I^{c_1}_h + \Theta - \pi^2T_0] = \frac{g^2}{16\pi^2}C_F[-2\ln\lambda^2a^2 + c],
\end{equation} 
where
\begin{equation}\label{eq:c_results}
c = \left\{
\begin{array}{ll}
 4.525 & {\rm Wilson, c_1 = 0} \\
 1.905 & {\rm Iwasaki, c_1 = -0.331} \\
 -2.495 & {\rm DBW2, c_1 = -1.40686} \\
 3.930 & {\rm Symanzik, c_1 = -1/12 }
\end{array}
\right.,
\end{equation}
with a maximum error of $\pm 1$ in the last digit. Our result for the Wilson case agrees with the result of Ref.~\cite{Flynn:1990qz}.

\subsection{$B_B$ renormalization}
In QCD, the B-meson parameter $B_B$ is defined as
\begin{equation}
B_B \equiv \frac{\langle \bar{B} | O^{\rm \overline{MS}(NDR)}_{VV+AA} | B \rangle}{\frac{8}{3} f^{2}_B m^2_B}.
\end{equation}
Combining the results of this section with the results of Sec.~\ref{sec:match_full_effective_op} and Sec.~\ref{sec:match_effective_lattice_op} we can relate $B_B$ to lattice as 
\begin{equation}\label{eq:b_b_mix}
B_B = \frac{Z_1 Z_{L}}{C^2_{A_0} Z^2_A} B^{\rm lat}_{VV+AA} + \frac{Z_2 Z_{S}}{C^2_{A_0} Z^2_A} B^{\rm lat}_{SS+PP} ,
\end{equation}
where 
\begin{equation}\label{eq:b_latt_def}
B^{\rm lat}_{O_i} \equiv \frac{\langle \bar{B}|O^{\rm lat}_{i}(0)|B\rangle }{\frac{8}{3}|\langle 0|A^{\rm lat}_0|B\rangle|^2m_B} .
\end{equation}
Note that the $B_B$ renormalization does not depend on $(1-w^2_0)^{-1}Z^{-1}_w$ since it cancels in the ratios.

\section{Mean-field improvement}
\label{sec:mean_field_impr}
Finally, let us briefly discuss the application of mean-field improvement to our results. We follow the much more detailed discussion in Refs.~\cite{Aoki:1998vv,Aoki:2002iq,Hernandez:1994bx}. The main reason for the improvement program is the fact that for some values of $M_5$, the DWF-specific factor $Z_w$ becomes quite large \cite{Aoki:1997xg}. This problem can be circumvented by substituting $w_0$ and $Z_w$ with:
\begin{equation}
w^{\rm MF}_0 = w_0 + 4(1-u),
\end{equation}
\begin{equation}
Z^{\rm MF}_w = Z_w|_{w_0 = w^{\rm MF}_0} + \frac{4w^{\rm MF}_0}{1-(w^{\rm MF}_0)^2}2(1-u),
\end{equation}
where $u = P^{1/4}$, with $P$ being the value of the plaquette. 

The mean field improvement will also affect other perturbative constants. The constant $f=1/2-z_2$ in Eqs.~(\ref{eq:Za_final},\ref{eq:d_l}) should be replaced by $f^{\rm MF} = 1/2-z^{\rm MF}_2$, with $z^{\rm MF}_2$ of Refs.~\cite{Aoki:1998vv,Aoki:2002iq}.
Because of the change in $w_0$, the value of the constant $d$ in Eq.~(\ref{eq:vertex_result}) should also change. One can obtain the values of $d^{\rm MF}$ from $d$ by observing that $d^{\rm MF}(M_5) = d(\tilde{M}_5)$, where $\tilde{M}_5 = M_5 -4(1-u)$. Similarly, the constant $v$ in Eq.~(\ref{eq:d_l}) should be replaced by $v^{\rm MF}(M_5) = v(\tilde{M}_5)$.  
For the heavy quark, mean-field improvement has no effect on the wave function renormalization as was pointed out in Refs.~\cite{Bernard:1993fq,Hernandez:1994bx}. However, one should be mindful of the subtlety involving different normalization conditions for $Z^h_2$ discussed in Refs.~\cite{Eichten:1989kb,Boucaud:1992nf}. The normalization condition consistent with the mean-field improvement necessitates the use of $e-\delta M$ instead of $e$ in Eqs.~(\ref{eq:Za_final},\ref{eq:d_l}). The constant $c$ in Eq.~(\ref{eq:c_results}) is unaffected by mean-field improvement. Finally, the light-quark field should be multiplied by $\sqrt{u}$, which leads to the multiplication of $Z_{\Gamma}$ by $\sqrt{u}$ and of $Z_{L,S}$ by $u$.  
\section{Conclusions}
\label{sec:conclusions}
In this paper we have calculated one-loop renormalization constants for operators combining static heavy and domain-wall light quarks. We obtained results for Wilson, Iwasaki, DBW2 and Symanzik gauge actions.
We have confirmed that all bilinears regularize by a single constant when matched to the static theory.  We have shown that the combination of domain-wall fermions and an improved gauge action reduces the size of perturbative corrections significantly compared to the Wilson case. This suggests that our choice of action is quite suitable for the determination of $f_B$, $B_B$ and $\Delta m_s/\Delta m_d$ with increased precision. Our results provide the necessary connection between those phenomenological quantities and lattice calculations.

\section*{Acknowledgments}
We would like to thank N. Christ, S. Aoki and S. D. Cohen for useful discussions. We are grateful to Y. Taniguchi for the copy of his notes on DWF perturbation theory. This work was supported in part by the U.S. Department of Energy under grant No. DE-FG02-92ER40699, by the Grant-in-Aid of the Ministry of Education, Culture, Sports, Science and Technology, Japan (MEXT Grant), No. 17740138.
                                                                                  
\section{Appendix}
The improved gauge propagator in momentum space is given by \cite{Weisz:1983bn}
\begin{equation}\label{eq:gauge_prop_gen}
D^{c_1}_{\mu\nu}(k) = \frac{1}{(\widehat{k}^2)^2}\left(\big[\alpha-A^{c_1}_{\mu\nu}(k)\big] \widehat{k}_\mu\widehat{k}_\nu + \delta_{\mu\nu}\sum_\sigma \widehat{k}^2_\sigma A^{c_1}_{\nu\sigma}(k) \right),
\end{equation}
where $\alpha$ is the gauge parameter, $\widehat{k}_\mu = 2\sin(k_\mu/2)$, $\widehat{k}^2 = 4 \sum_\mu \sin^2(k_\mu/2)$  and
\begin{multline}
A^{c_1}_{\mu\nu} (k) = A^{c_1}_{\nu\mu} (k) = \frac{1-\delta_{\mu\nu}}{\Delta (k)} \,
\Big[ (\widehat{k}^2)^2
-c_1 \widehat{k}^2 \Big( 2\sum_\rho \widehat{k}_\rho^4
+\widehat{k}^2 \sum_{\rho \neq \mu,\nu} \widehat{k}_\rho ^2 \Big) \\
+ c_1^2 \Big( \Big( \sum_\rho \widehat{k}_\rho^4 \Big)^2
+\widehat{k}^2 \sum_\rho \widehat{k}_\rho^4 \sum_{\tau \neq \mu,\nu}
\widehat{k}_\tau^2 + (\widehat{k}^2)^2 \prod_{\rho \neq \mu,\nu}
\widehat{k}_\rho^2 \Big) \Big] ,
\end{multline}
with
\begin{multline}
\Delta (k) = \Big( \widehat{k}^2 -c_1 \sum_\rho \widehat{k}_\rho^4 \Big)
\Big[ \widehat{k}^2
-c_1 \Big( (\widehat{k}^2)^2 + \sum_\tau \widehat{k}_\tau^4 \Big)
+\frac{1}{2} c_1^2 \Big( (\widehat{k}^2)^3 + 2 \sum_\tau \widehat{k}_\tau^6
- \widehat{k}^2 \sum_\tau \widehat{k}_\tau^4 \Big) \Big] \\
-4 c_1^3 \sum_\rho \widehat{k}_\rho^4 \prod_{\tau \neq \rho}
\widehat{k}_\tau^2 .
\end{multline}
The different choices for the parameter $c_1$ correspond to different gauge actions discussed in the main text of this paper. In the case $c_1 = 0$, the propagator in Eq.~(\ref{eq:gauge_prop_gen}) reduces to the standard Wilson plaquette action propagator.


\bibliography{paper}


\begin{table}[htbp]
\caption{\label{tab:dwf_tad}Results for vertex integrals defined in Sec.~\ref{sec:match_effective_lattice_op}.}
\begin{ruledtabular}
\begin{center}
\begin{tabular}{c c c c}
$M_5$ & $I_w$ & ${\mathcal I}_\chi^w$ &$\hat{\mathcal I}_\chi^w$\\
\hline                  
$0.05$ & $-4.138229(319)$ & $-1.909304(325)$ & $-0.703771(17)$ \\ 
$0.10$ & $-4.034432(236)$ & $-2.030270(283)$ & $-0.712970(18)$ \\ 
$0.15$ & $-3.948298(192)$ & $-2.133953(315)$ & $-0.722356(18)$ \\ 
$0.20$ & $-3.872189(163)$ & $-2.228815(258)$ & $-0.731938(18)$ \\ 
$0.25$ & $-3.802874(143)$ & $-2.316957(249)$ & $-0.741722(18)$ \\ 
$0.30$ & $-3.738464(129)$ & $-2.400620(227)$ & $-0.751716(18)$ \\ 
$0.35$ & $-3.677754(117)$ & $-2.480180(264)$ & $-0.761929(19)$ \\ 
$0.40$ & $-3.619923(108)$ & $-2.558862(216)$ & $-0.772367(19)$ \\ 
$0.45$ & $-3.564378(100)$ & $-2.634128(243)$ & $-0.783041(19)$ \\ 
$0.50$ & $-3.510664(93)$ & $-2.709506(214)$ & $-0.793959(19)$ \\ 
$0.55$ & $-3.458427(87)$ & $-2.782736(229)$ & $-0.805132(20)$ \\ 
$0.60$ & $-3.407373(81)$ & $-2.855975(230)$ & $-0.816570(20)$ \\ 
$0.65$ & $-3.357255(75)$ & $-2.928999(222)$ & $-0.828285(20)$ \\ 
$0.70$ & $-3.307856(70)$ & $-3.001320(248)$ & $-0.840288(20)$ \\ 
$0.75$ & $-3.258987(65)$ & $-3.074116(255)$ & $-0.852592(20)$ \\ 
$0.80$ & $-3.210479(60)$ & $-3.147106(260)$ & $-0.865211(21)$ \\ 
$0.85$ & $-3.162165(55)$ & $-3.220438(268)$ & $-0.878159(21)$ \\ 
$0.90$ & $-3.113899(50)$ & $-3.294598(273)$ & $-0.891453(21)$ \\ 
$0.95$ & $-3.065532(44)$ & $-3.369698(280)$ & $-0.905107(21)$ \\ 
$1.00$ & $-3.016959(36)$ & $-3.445750(258)$ & $-0.919141(22)$ \\ 
$1.05$ & $-2.967904(40)$ & $-3.523295(291)$ & $-0.933573(22)$ \\ 
$1.10$ & $-2.918434(44)$ & $-3.602252(251)$ & $-0.948424(22)$ \\ 
$1.15$ & $-2.868167(47)$ & $-3.682678(253)$ & $-0.963696(24)$ \\ 
$1.20$ & $-2.816869(48)$ & $-3.764914(309)$ & $-0.979474(23)$ \\ 
$1.25$ & $-2.764631(52)$ & $-3.849567(315)$ & $-0.995722(23)$ \\ 
$1.30$ & $-2.711104(56)$ & $-3.936509(322)$ & $-1.012510(29)$ \\ 
$1.35$ & $-2.656052(60)$ & $-4.026280(329)$ & $-1.029829(30)$ \\ 
$1.40$ & $-2.599188(65)$ & $-4.119041(336)$ & $-1.047733(30)$ \\ 
$1.45$ & $-2.540188(72)$ & $-4.215449(345)$ & $-1.066257(31)$ \\ 
$1.50$ & $-2.478673(79)$ & $-4.315845(352)$ & $-1.085443(31)$ \\ 
$1.55$ & $-2.414181(87)$ & $-4.421470(280)$ & $-1.105334(32)$ \\ 
$1.60$ & $-2.346139(97)$ & $-4.531634(286)$ & $-1.125985(35)$ \\ 
$1.65$ & $-2.273804(108)$ & $-4.647352(381)$ & $-1.147436(33)$ \\ 
$1.70$ & $-2.196232(120)$ & $-4.770915(392)$ & $-1.169762(34)$ \\ 
$1.75$ & $-2.112133(135)$ & $-4.903247(399)$ & $-1.193028(34)$ \\ 
$1.80$ & $-2.019661(155)$ & $-5.046261(405)$ & $-1.217309(35)$ \\ 
$1.85$ & $-1.916047(182)$ & $-5.202899(419)$ & $-1.242693(36)$ \\ 
$1.90$ & $-1.796524(223)$ & $-5.378317(435)$ & $-1.269279(37)$ \\ 
$1.95$ & $-1.651230(292)$ & $-5.583553(420)$ & $-1.297141(31)$ \\ 

\end{tabular}
\end{center}
\end{ruledtabular}
\end{table}    

\begin{table}[htbp]
\caption{\label{tab:dwf_i-w}Results for $I^{c_1}_\chi$ integral defined in Eq.~(\ref{eq:I_chi_def_diff}).}
\begin{ruledtabular}
\begin{center}
\begin{tabular}{c c c c}
$M_5$ & Iwasaki & DBW2 & Symanzik \\
\hline                  
0.05  & 0.246331(9) & 0.408168(15) & 0.100822(4) \\ 
0.10  & 0.249861(9) & 0.414136(16) & 0.102247(4) \\ 
0.15  & 0.253496(9) & 0.420275(16) & 0.103714(4) \\ 
0.20  & 0.257237(10) & 0.426590(16) & 0.105225(4) \\ 
0.25  & 0.261091(10) & 0.433088(16) & 0.106782(4) \\ 
0.30  & 0.265062(10) & 0.439774(17) & 0.108386(4) \\ 
0.35  & 0.269153(10) & 0.446657(17) & 0.110041(4) \\ 
0.40  & 0.273371(10) & 0.453742(17) & 0.111747(4) \\ 
0.45  & 0.277720(10) & 0.461039(18) & 0.113507(4) \\ 
0.50  & 0.282207(11) & 0.468556(18) & 0.115323(4) \\ 
0.55  & 0.286837(11) & 0.476302(18) & 0.117199(5) \\ 
0.60  & 0.291617(11) & 0.484287(18) & 0.119138(5) \\ 
0.65  & 0.296554(11) & 0.492522(19) & 0.121141(5) \\ 
0.70  & 0.301655(11) & 0.501018(19) & 0.123212(5) \\ 
0.75  & 0.306929(12) & 0.509787(19) & 0.125355(5) \\ 
0.80  & 0.312383(12) & 0.518844(20) & 0.127573(5) \\ 
0.85  & 0.318029(12) & 0.528201(20) & 0.129871(5) \\ 
0.90  & 0.323874(12) & 0.537875(21) & 0.132253(5) \\ 
0.95  & 0.329931(13) & 0.547883(21) & 0.134722(5) \\ 
1.00  & 0.336211(13) & 0.558242(21) & 0.137285(5) \\ 
1.05  & 0.342726(13) & 0.568972(22) & 0.139946(6) \\ 
1.10  & 0.349490(13) & 0.580095(22) & 0.142712(6) \\ 
1.15  & 0.356519(14) & 0.591634(23) & 0.145589(6) \\ 
1.20  & 0.363828(14) & 0.603614(23) & 0.148583(6) \\ 
1.25  & 0.371435(14) & 0.616064(24) & 0.151703(6) \\ 
1.30  & 0.379360(15) & 0.629014(24) & 0.154956(6) \\ 
1.35  & 0.387625(15) & 0.642497(25) & 0.158353(6) \\ 
1.40  & 0.396252(16) & 0.656551(25) & 0.161902(7) \\ 
1.45  & 0.405268(16) & 0.671216(26) & 0.165615(7) \\ 
1.50  & 0.414702(16) & 0.686539(27) & 0.169504(7) \\ 
1.55  & 0.424586(17) & 0.702570(27) & 0.173584(7) \\ 
1.60  & 0.434956(17) & 0.719365(28) & 0.177868(8) \\ 
1.65  & 0.445852(18) & 0.736989(29) & 0.182376(8) \\ 
1.70  & 0.457321(19) & 0.755513(30) & 0.187125(8) \\ 
1.75  & 0.469413(19) & 0.775019(31) & 0.192138(8) \\ 
1.80  & 0.482186(20) & 0.795598(32) & 0.197440(9) \\ 
1.85  & 0.495709(21) & 0.817357(33) & 0.203059(9) \\ 
1.90  & 0.510059(22) & 0.840417(35) & 0.209029(10) \\ 
1.95  & 0.525326(24) & 0.864919(37) & 0.215387(10) \\ 

\end{tabular}
\end{center}
\end{ruledtabular}
\end{table}    

\begin{table}[htbp]
\caption{\label{tab:vertex}Results for $d$ defined in Eq.~(\ref{eq:vertex_result}).}
\begin{ruledtabular}
\begin{center}
\begin{tabular}{c c c c c}
$M_5$ & Wilson & Iwasaki & DBW2 & Symanzik \\
\hline                  
0.05   & 5.343762(661) & 5.097431(670) & 4.935594(676) & 5.242940(665) \\ 
0.10   & 5.351732(537) & 5.101871(546) & 4.937596(553) & 5.249485(541) \\ 
0.15   & 5.359895(525) & 5.106399(534) & 4.939620(541) & 5.256181(529) \\ 
0.20   & 5.369066(439) & 5.111829(449) & 4.942476(455) & 5.263841(443) \\ 
0.25   & 5.378109(410) & 5.117018(420) & 4.945021(426) & 5.271327(414) \\ 
0.30   & 5.387368(374) & 5.122306(384) & 4.947594(391) & 5.278982(378) \\ 
0.35   & 5.396005(400) & 5.126852(410) & 4.949348(417) & 5.285964(404) \\ 
0.40   & 5.406418(343) & 5.133047(353) & 4.952676(360) & 5.294671(347) \\ 
0.45   & 5.415465(362) & 5.137745(372) & 4.954426(380) & 5.301958(366) \\ 
0.50   & 5.426211(326) & 5.144004(337) & 4.957655(344) & 5.310888(330) \\ 
0.55   & 5.436031(336) & 5.149194(347) & 4.959729(354) & 5.318832(341) \\ 
0.60   & 5.446778(331) & 5.155161(342) & 4.962491(349) & 5.327640(336) \\ 
0.65   & 5.457969(317) & 5.161415(328) & 4.965447(336) & 5.336828(322) \\ 
0.70   & 5.468888(338) & 5.167233(349) & 4.967870(357) & 5.345676(343) \\ 
0.75   & 5.480511(340) & 5.173582(352) & 4.970724(359) & 5.355156(345) \\ 
0.80   & 5.492374(341) & 5.179991(353) & 4.973530(361) & 5.364801(346) \\ 
0.85   & 5.504444(344) & 5.186415(356) & 4.976243(364) & 5.374573(349) \\ 
0.90   & 5.517044(344) & 5.193170(356) & 4.979169(365) & 5.384791(349) \\ 
0.95   & 5.530123(345) & 5.200192(358) & 4.982240(366) & 5.395401(350) \\ 
1.00   & 5.543568(316) & 5.207357(329) & 4.985326(337) & 5.406283(321) \\ 
1.05   & 5.557626(353) & 5.214900(366) & 4.988654(375) & 5.417680(359) \\ 
1.10   & 5.572262(317) & 5.222772(330) & 4.992167(339) & 5.429550(323) \\ 
1.15   & 5.587149(324) & 5.230630(338) & 4.995515(347) & 5.441560(330) \\ 
1.20   & 5.602309(380) & 5.238481(394) & 4.998695(403) & 5.453726(386) \\ 
1.25   & 5.618476(390) & 5.247041(404) & 5.002412(414) & 5.466773(396) \\ 
1.30   & 5.635103(407) & 5.255743(422) & 5.006089(431) & 5.480147(413) \\ 
1.35   & 5.652503(419) & 5.264878(434) & 5.010006(444) & 5.494150(425) \\ 
1.40   & 5.670496(431) & 5.274244(447) & 5.013945(456) & 5.508594(438) \\ 
1.45   & 5.689380(448) & 5.284112(464) & 5.018164(474) & 5.523765(455) \\ 
1.50   & 5.709075(462) & 5.294373(478) & 5.022536(489) & 5.539571(469) \\ 
1.55   & 5.730317(399) & 5.305731(416) & 5.027747(426) & 5.556733(406) \\ 
1.60   & 5.751788(418) & 5.316832(435) & 5.032423(446) & 5.573920(426) \\ 
1.65   & 5.773720(522) & 5.327868(540) & 5.036731(551) & 5.591344(530) \\ 
1.70   & 5.797385(546) & 5.340064(565) & 5.041872(576) & 5.610260(554) \\ 
1.75   & 5.822352(568) & 5.352939(587) & 5.047333(599) & 5.630214(576) \\ 
1.80   & 5.848613(595) & 5.366427(615) & 5.053015(627) & 5.651173(604) \\ 
1.85   & 5.876253(637) & 5.380544(658) & 5.058896(670) & 5.673194(646) \\ 
1.90   & 5.905562(695) & 5.395503(717) & 5.065145(730) & 5.696533(705) \\ 
1.95   & 5.937642(743) & 5.412316(767) & 5.072723(780) & 5.722255(753) \\ 

\end{tabular}
\end{center}
\end{ruledtabular}
\end{table}


\end{document}